\documentclass{article}%
\usepackage{amsmath}
\usepackage{graphicx}%
\usepackage{amsfonts}%
\usepackage{amssymb}

\begin{document}

\title{{\Large Constraint Reorganization Consistent with the Dirac Procedure}}
\author{D.M. Gitman\thanks{Institute of Physics, University of Sao Paulo, Brazil;
e-mail: gitman@fma.if.usp.br} \ and I.V. Tyutin\thanks{Lebedev Physics
Institute, Moscow, Russia; e-mail: tyutin@lpi.ru}}
\date{\today }
\maketitle

\begin{abstract}
The way of finding all the constraints in the Hamiltonian formulation of
singular (in particular, gauge) theories is called the Dirac procedure. The
constraints are naturally classified according to the correspondig stages of
this procedure. On the other hand, it is convenient to reorganize the
constraints such that they are explicitly decomposed into the first-class and
second-class constraints. The presence of the first-class constraints is
related to the existence of gauge symmetries in the theory. The second-class
constraints can be used to formulate the equations of motion and the
quantization procedure in an invariant form by means of the Dirac brackets. We
discuss the reorganization of the constraints into the first- and second-class
constraints that is consistent with the Dirac procedure, i.e., that does not
violate the decomposition of the constraints according to the stages of the
Dirac procedure. The possibility of such a reorganization is important for the
study of gauge symmetries in the Lagrangian and Hamiltonian formulations.

\end{abstract}

\section{Introduction}

It is well known that from the Hamiltonian formulation standpoint, almost all
modern physical theories are theories with constraints
\cite{Dirac64,GitTy90,HenTe92}.\ An information about the constraint structure
is important for the physical sector identification, for the study of
classical and quantum symmetries, for quantization purposes and so on.

The complete set of constraints in the Hamiltonian formulation defines \ a
constraint surface where the dynamics evolves. To describe this surface one
can use different sets of equivalent constraints. We call the passage from
some set of constraints to another equivalent one the reorganization of
constraints. Dirac remarked that it is convenient to reorganize the
constraints such that they explicitly \ split into the\ first-class
constraints (FCC) and the\ second-class constraints (SCC). The presence of FCC
is related to the existence of gauge symmetries in the theory. SCC can be used
to formulate the\ equations of motion and the quantization procedure in an
invariant form by means of the Dirac brackets. The way of finding all the
constraints in the Hamiltonian formulation is usually called the Dirac
procedure (DP). We recall that after the Hamiltonization, a singular
Lagrangian theory is described by the Hamilton equations of motion with the
primary constraints \cite{Dirac64,GitTy90,HenTe92},
\begin{equation}
\dot{\eta}=\left\{  \eta,H^{\left(  1\right)  }\right\}  ,\;\Phi^{\left(
1\right)  }\left(  \eta\right)  =0,\;H^{\left(  1\right)  }=H\left(
\eta\right)  +\lambda\Phi^{\left(  1\right)  }\left(  \eta\right)
,\;\dot{\eta}\equiv\frac{d\eta}{dt}.\label{1.1}%
\end{equation}
Here $\eta=\left(  q,p\right)  $ are phase-space variables; $\Phi^{\left(
1\right)  }\left(  \eta\right)  =0$ are the primary constraints (we suppose
that all the primary constraints are independent); $\lambda$'s are the
Lagrange multipliers to the primary constraints; $H=H\left(  \eta\right)  $ is
the Hamiltonian, and $H^{\left(  1\right)  }$ is the total Hamiltonian.
$\{F,G\}$ denotes the Poisson bracket of two functions $F\left(  \eta\right)
$ and $G\left(  \eta\right)  $. Sometimes, the additional variables $\lambda
$'s can be partially or completely eliminated from Eqs. (\ref{1.1}). Moreover,
some new constraints (additional to the primary ones) may exist in the theory.
The way of eliminating $\lambda$'s and finding new constraints was proposed by
Dirac \cite{Dirac64} and, as was already said, is called DP. DP is a part of
the complete Hamiltonization of a singular Lagrange theory. The procedure is
based on the so called consistency conditions $\dot{\Phi}=0$ which have to
hold for any constraint equation $\Phi=0$. In the general case, the
Hamiltonian $H$ and the constraints\footnote{We call often the functions
$\Phi$ constraints as well.} $\Phi$ may depend on time $t$ explicitly. We take
such a possibility into account . However, the argument $t$ will not be
written explicitly. Using Eqs. (\ref{1.1}), we can transform the consistency
condition to the form
\begin{equation}
\dot{\Phi}=\frac{\partial\Phi}{\partial t}+\frac{\partial\Phi}{\partial\eta
}\dot{\eta}=\left\{  \Phi\,,H^{\left(  1\right)  }+\epsilon\right\}
=0.\label{1.2}%
\end{equation}
Here, $\epsilon$ is the momentum conjugate to time $t$ and the Poisson
brackets are defined in the extended phase space of the variables
$\eta;t,\epsilon$, see for details \cite{GitTy90}. Finding the primary
constraints can be considered the first stage of DP. At the second stage of
DP, we apply the consistency conditions (\ref{1.2}) to the primary constraints
trying to define some $\lambda$'s. Those $\lambda$'s that can be defined here
are denoted by $\lambda_{1}=\bar{\lambda}_{1}\left(  \eta\right)  $. In
addition, we can reveal some new independent constraints$\ \Phi^{\left(
2\right)  }=0$; we call them the second-stage constraints. We can substitute
expressions for $\lambda_{1}$ directly in the total Hamiltonian to construct
the Hamiltonian $H_{1}^{\left(  1\right)  }=\left.  H^{\left(  1\right)
}\right|  _{\lambda_{1}=\bar{\lambda}_{1}}\,.$ At the third stage, we use the
consistency conditions for the second-stage constraints to find some $\lambda
$'s (these are denoted by $\lambda_{2}=\bar{\lambda}_{2}\left(  \eta\right)
$) and to reveal some new third-stage\emph{\ }constraints $\ \Phi^{\left(
3\right)  }\left(  \eta\right)  =0$ independent from the previous ones. We can
substitute expressions for $\lambda_{(2)}$ directly in the Hamiltonian
$H^{\left(  1\right)  }$ to construct the Hamiltonian $H_{2}^{\left(
1\right)  }=\left.  H^{\left(  1\right)  }\right|  _{\lambda_{1}=\bar{\lambda
}_{1},\lambda_{2}=\bar{\lambda}_{2}}\,.$ Continuing DP, we can determine some
$\lambda$'s and obtain some new independent constraints. At the $r$-th stage,
we obtain $\lambda_{r-1}=\bar{\lambda}_{r-1}\left(  \eta\right)  $ and
$\Phi^{\left(  r\right)  }\left(  \eta\right)  =0$ and construct the
Hamiltonian $H_{r-1}^{\left(  1\right)  }=\left.  H^{\left(  1\right)
}\right|  _{\lambda_{1}=\bar{\lambda}_{1},...,\lambda_{r-1}=\bar{\lambda
}_{r-1}}\,.$ Because the number of the degrees of freedom is finite and EM are
assumed to be consistent, DP stops at a certain $k$-th stage, after which new
constraints do not appear. We will refer to all the constraints that were
obtained by DP and differ from the primary constraints as the secondary
constraints, that is, the secondary constraints are the second-, third-, and
etc. stage constraints.

We use the notation $\Phi^{\left(  i,...,j\right)  }\equiv\left(
\Phi^{\left(  i\right)  },...,\Phi^{\left(  j\right)  }\right)  \,,\;1\leq
i<j\leq k\,.$ The secondary constraints are then $\Phi^{\left(
2,...,k\right)  }$, and the complete set of constraints, both primary and
secondary, of the theory is $\Phi=\Phi^{\left(  1,...,k\right)  }=\left(
\Phi^{\left(  1\right)  },...,\Phi^{\left(  k\right)  }\right)  \,.$
Schematically, DP can be represented as
\begin{equation}
\Phi^{\left(  1\right)  }\rightarrow%
\begin{array}
[c]{l}%
\\
\Phi^{\left(  2\right)  }\\
\bar{\lambda}_{1}%
\end{array}
\rightarrow%
\begin{array}
[c]{l}%
\\
\Phi^{\left(  3\right)  }\\
\bar{\lambda}_{2}%
\end{array}
\rightarrow\cdots\rightarrow%
\begin{array}
[c]{l}%
\\
\Phi^{\left(  k\right)  }\\
\bar{\lambda}_{k-1}%
\end{array}
\rightarrow%
\begin{array}
[c]{l}%
\\
O\left(  \Phi\right) \\
\bar{\lambda}_{k}%
\end{array}
\,.\label{1.3}%
\end{equation}
Here, the arrow $\,\rightarrow\,$ implies DP, and $O(\Phi)$ denotes the terms
proportional to the functions $\Phi,$ i.e., $\left.  O(\Phi)\right|  _{\Phi=0}=0\,.$

The important question is: does a consistent with DP constraint reorganization
to the first- and second-class constraints exist, i.e., the reorganization
that does not violate the decomposition of the constraints\ according to their
stages in DP. The problem is important for understanding \ the general
structure of singular theories. In particular, the existence of such a
reorganization is a crucial point for finding a relation between the
constraint structure and the symmetry structure of singular (gauge) theories
\cite{BorTy98}. This problem was considered by many authors \cite{Others}.
However, in these publications, either the theories of a particular form were
considered or too restrictive assumptions were used.

In the present paper, we are going to demonstrate that a complete set of
constraints $\Phi$ can be reorganized to the chains of SCC and FCC in
consistency with their hierarchy in DP. The possibility of such a constraint
reorganization is formulated as the following statement.

It is possible to reorganize the complete set of constraints obtained in DP to
the form:\ $\Phi=\left(  \Phi^{\left(  i\right)  }\right)  =0\,,\;\Phi
^{\left(  i\right)  }=(\varphi^{(i)};\chi^{\left(  i\right)  }%
),\;i=1,...,k,\;(\chi^{(k)}\equiv0),$ where $\Phi^{\left(  i\right)  }$ are
the constraints of the $i$-th stage, $\varphi^{(i)}$ are the SCC functions\ of
the $i$-th stage, $\chi^{\left(  i\right)  }$ are the FCC functions of the
$i$-th stage, and $k$ is the number of all the stages of DP.

The constraints $\chi^{\left(  i\right)  }$ and $\varphi^{(i)}$ are decomposed
into the groups
\begin{equation}
\varphi^{\left(  i\right)  }=\left(  \varphi^{\left(  i|u\right)
},\;u=1,...,k\right)  ,\;\;\chi^{\left(  i\right)  }=\left(  \chi^{\left(
i|a\right)  },\;a=1,...,k-1\right) \label{1.4}%
\end{equation}
such that the total Hamiltonian and the Lagrange multipliers $\lambda$ have
the form
\begin{equation}
H^{\left(  1\right)  }=H+\lambda_{\varphi^{u}}\varphi^{\left(  1|u\right)
}+\lambda_{\chi^{a}}\chi^{\left(  1|a\right)  }\,,\;\;\lambda=\left(
\lambda_{\varphi^{u}},\lambda_{\chi^{a}}\right)  \,.\label{1.5}%
\end{equation}
Each of the groups $\varphi^{\left(  i|u\right)  }$, $\chi^{\left(
i|a\right)  },\lambda_{\varphi^{u}}$, and $\lambda_{\chi^{a}}$ may be either
empty or contain several functions: $\varphi^{\left(  i|u\right)  }=\left(
\varphi_{\mu_{u}}^{\left(  i|u\right)  }\,,\;\mu_{u}=1,...,r_{u}\right)
,\;$\ $\;\chi^{\left(  i|a\right)  }=\left(  \chi_{\rho_{a}}^{\left(
i|a\right)  }\,,\;\rho_{a}=1,...,s_{a}\right)  ,$ $\;\lambda_{\varphi^{u}%
}=\left(  \lambda_{\varphi^{u}}^{\mu_{u}}\right)  ,$ $\lambda_{\chi^{a}%
}=(\lambda_{\chi^{a}}^{\rho_{a}}).$ Each group $\varphi^{\left(  1|u\right)
}$ and $\;\chi^{\left(  1|a\right)  }\;$produces a chain of the groups of
constraints of the second, third, and so on stages within DP,\ $\varphi
^{\left(  1|u\right)  }\rightarrow\varphi^{\left(  2|u\right)  }%
\rightarrow\varphi^{\left(  3|u\right)  }\rightarrow\cdot\cdot\cdot
\rightarrow\varphi^{\left(  u|u\right)  }\rightarrow\lambda_{\varphi^{u}}%
=\bar{\lambda}_{u},$ $\;\chi^{\left(  1|a\right)  }\rightarrow\chi^{\left(
2|a\right)  }\rightarrow\chi^{\left(  3|a\right)  }\rightarrow\cdot\cdot
\cdot\rightarrow\chi^{\left(  a|a\right)  }$. Here, the indices $u$ and $a$
after the sign of the vertical bar in the superscripts number the constraint
chains. All the constraints in a chain are of the same class, and all the
groups in the chain have the same number of constraints.

The chain of SCC with the number $u$ ends up with the group of the
$u$-th-stage constraints. Their consistency conditions define\ the
$\lambda_{\varphi^{u}}-$ multipliers. The chain of FCC with the number $a$
ends up with the group of the $a$-th-stage constraints. The\ constraints of
the last group of any chain of FCC are not involved in determining new constraints.

The Poisson brackets of the SCC constraints from different chains vanish on
the constraint surface.

The described hierarchy of the constraints in DP schematically looks as
follows:
\[
\begin{array}
[c]{llllllllll}%
\varphi^{\left(  1|1\right)  } & \rightarrow & \bar{\lambda}_{1} &  &  &  &  &
&  & \\
\varphi^{\left(  1|2\right)  } & \rightarrow & \varphi^{\left(  2|2\right)  }
& \rightarrow & \bar{\lambda}_{2} &  &  &  &  & \\
\vdots & \vdots & \vdots & \vdots & \vdots & \vdots &  &  &  & \\
\varphi^{\left(  1|k-1\right)  } & \rightarrow & \varphi^{\left(
2|k-1\right)  } &  & \;\;\cdots\rightarrow & \varphi^{\left(  k-1|k-1\right)
} & \rightarrow & \bar{\lambda}_{k-1} &  & \\
\varphi^{\left(  1|k\right)  } & \rightarrow & \varphi^{\left(  2|k\right)  }
&  & \;\;\cdots\rightarrow & \varphi^{\left(  k-1|k\right)  } & \rightarrow &
\varphi^{\left(  k|k\right)  } & \rightarrow & \bar{\lambda}_{k}\\
\chi^{\left(  1|k-1\right)  } & \rightarrow & \chi^{\left(  2|k-1\right)  } &
& \;\;\cdots\rightarrow & \chi^{\left(  k-1|k-1\right)  } & \rightarrow &
O(\Phi) &  & \\
\vdots & \vdots & \vdots & \vdots & \vdots & \vdots &  &  &  & \\
\chi^{\left(  1|2\right)  } & \rightarrow & \chi^{\left(  2|2\right)  } &
\rightarrow & O(\Phi^{\left(  1,2,3\right)  }) &  &  &  &  & \\
\chi^{\left(  1|1\right)  } & \rightarrow & O(\Phi^{\left(  1,2\right)  }) &
&  &  &  &  &  &
\end{array}
.
\]
In addition\footnote{Here and in what follows, we use the notation
\par
$\left[  f\right]  =\mathrm{the\;number\;of\;the\;functions\;}f,\;$%
i.e.,\ $f=(f_{i},\;i=1,...,n)\rightarrow\;\left[  f\right]  =n.$}:%
\begin{align*}
& \left[  \lambda_{\varphi^{u}}\right]  =[\bar{\lambda}_{u}]=[\varphi^{\left(
i|u\right)  }]=[\varphi^{\left(  1|u\right)  }],\quad\left[  \lambda_{\chi
^{a}}\right]  =[\chi^{\left(  i|a\right)  }]=[\chi^{\left(  1|a\right)
}]\,,\\
& \,\{\varphi^{\left(  i|u\right)  },\varphi^{\left(  j|v\right)  }%
\}=\{\Phi\},\quad\{\chi^{\left(  i|a\right)  },\Phi\}=O(\Phi),\;u\neq v\,,\\
& u,v=1,...,k\,,\;a=1,...,k-1\,.
\end{align*}

In what follows, we present a constructive proof of the statement. Namely,
considering a specific version of DP (which we call the refined DP), we
construct the above-mentioned set of constraints. It is this set of
constraints that we call the constraints consistent with DP.

\section{Refined Dirac procedure}

We begin with some remarks about the\ theories under consideration. The only
restrictions to be imposed on the theories follow from the requirement of
\ applicability of DP. These requirements will be formulated in terms of the
ranks of some Jacobi matrices of the type $\partial\Phi^{\left(
1,...l\right)  }/\partial\eta$ and of the Poisson bracket matrices of the type
$\left\{  \Phi^{\left(  1\right)  },\Phi^{\left(  l\right)  }\right\}  .$ We
assume that these matrices are of a constant rank. Literally, this means that
they are of a constant rank in a vicinity of the point $\eta=0$ on the
corresponding constraint surface $\Phi^{\left(  1,...l\right)  }=0.$ We also
assume that $H=O\left(  \eta^{2}\right)  $ and $\Phi^{\left(  1\right)
}=O\left(  \eta\right)  .$ As was already said above, we assume that primary
constraint functions $\Phi^{\left(  1\right)  }$ are independent, that is,
\begin{equation}
\mathrm{rank}{\frac{\partial\Phi^{(1)}}{\partial\eta}}=[\Phi^{(1)}%
]=m_{1}\,.\label{2.1}%
\end{equation}

\subsection{First stage}

Consider the antisymmetric matrix $C_{\alpha_{0}\beta_{0}}^{(1)}%
=\{\Phi_{\alpha_{0}}^{(1)},\Phi_{\beta_{0}}^{(1)}\}.$ This matrix appears in
consistency conditions (\ref{1.2}) for the primary constraints,
\begin{equation}
\{\Phi_{\alpha_{0}}^{(1)},H^{\left(  1\right)  }+\epsilon\}=\{\Phi_{\alpha
_{0}}^{(1)},H+\epsilon\}+C_{\alpha_{0}\beta_{0}}^{(1)}\lambda^{\beta_{0}%
}=0\,.\label{2.1a}%
\end{equation}
We suppose that the matrix $C^{\left(  1\right)  }$ has a constant rank,
\ \textrm{rank\thinspace}$C^{\left(  1\right)  }=r_{1}\,$. Therefore, there
exists a submatrix of size $r_{1}\times r_{1}$, that is located on the
diagonal and is also antisymmetric, we let denote it by $M_{\mu_{1}\nu_{1}%
}^{\left(  1\right)  }\,,\;\left[  \nu_{1}\right]  =\left[  \mu_{1}\right]
=r_{1}\,,\;\det M^{\left(  1\right)  }\neq0.$ The equation
\[
C^{(1)}Z^{(1)}=O\left(  \Phi^{(1)}\right)
\]
has $m_{2}^{\prime}=m_{1}-r_{1}$ linearly independent solutions $Z_{\alpha
_{1}}^{(1)}$=$Z_{\alpha_{1}}^{(1)\alpha_{0}},\;[\alpha_{1}]=m_{2}^{\prime
}\,,\;$such that $\det Z_{\alpha_{1}}^{(1)\beta_{1}}\neq0$, $\alpha
_{0}=\left(  \mu_{1},\beta_{1}\right)  .$ Together with the vectors
$Z_{\mu_{1}}^{(1)}=Z_{\mu_{1}}^{(1)\alpha_{0}}=\delta_{\mu_{1}}^{\alpha_{0}},$
these solutions form a set of $m_{1}$ linearly independent vectors (see, for
example, \cite{GitTy90}, p.27). Using a nonsingular matrix $Z_{1},$ we
reorganize the primary constraints:
\begin{align*}
& \Phi^{(1)}\rightarrow Z_{1}\Phi^{(1)}=\left(
\begin{array}
[c]{l}%
\varphi_{\mu_{1}}^{\left(  1|1\right)  }=\Phi_{\mu_{1}}^{\left(  1\right)  }\\
\phi_{\alpha_{1}}^{(1|1)}=Z_{\alpha_{1}}^{(1)\alpha_{0}}\Phi_{\alpha_{0}%
}^{(1)}%
\end{array}
\right)  ,\;\alpha_{0}=\left(  \mu_{1},\alpha_{1}\right)  \,,\\
& Z_{1}=Z_{1\alpha_{0}}^{\beta_{0}}=\left(
\begin{array}
[c]{ll}%
\delta_{\mu_{1}}^{\nu_{1}} & 0\\
Z_{\alpha_{1}}^{(1)\nu_{1}} & Z_{\alpha_{1}}^{(1)\beta_{1}}%
\end{array}
\right)  ,\;\beta_{0}=\left(  \nu_{1},\beta_{1}\right)  \,.
\end{align*}
We thus have
\begin{equation}
\Phi^{(1)}\overset{Z}{\rightarrow}\left(
\begin{array}
[c]{l}%
\varphi^{\left(  1|1\right)  }\\
\phi^{(1|1)}%
\end{array}
\right)  \,,\;\;H^{\left(  1\right)  }=H+\lambda_{\varphi^{1}}\varphi^{\left(
1|1\right)  }+\lambda_{\phi^{1}}\phi^{(1|1)}\,.\label{2.2}%
\end{equation}
We call such a kind of reorganization the $Z-$reorganization.

We remark that any reorganization of primary constraints is always accompanied
by the corresponding $\lambda$-multiplier reorganization \cite{GitTy90}. These
new $\lambda$'s appear as the multipliers in front of the reorganized primary constraints.

The new primary constraints satisfy the properties
\begin{align}
& \left\{  \varphi^{\left(  1|1\right)  },\varphi^{\left(  1|1\right)
}\right\}  =M^{\left(  1\right)  }\,,\;\det M^{\left(  1\right)  }%
\neq0\,,\label{2.3}\\
& \left\{  \phi^{(1|1)},\varphi^{\left(  1|1\right)  }\right\}  =O\left(
\Phi^{(1)}\right)  ,\;\left\{  \phi^{(1|1)},\phi^{(1|1)}\right\}  =O\left(
\Phi^{(1)}\right)  \,.\label{2.4}%
\end{align}

Considering the consistency conditions (\ref{2.1a}) for the primary
constraints $\varphi^{\left(  1|1\right)  }$, we determine the Lagrange
multipliers $\lambda_{\varphi^{1}}:$%
\begin{equation}
\lambda_{\varphi^{1}}=\bar{\lambda}_{1}\equiv-\left[  M^{\left(  1\right)
}\right]  ^{-1}\left\{  \varphi^{\left(  1|1\right)  },\,H+\epsilon\right\}
\,.\label{2.5}%
\end{equation}
It is convenient to represent the Hamiltonian (\ref{2.2}) \ as
\begin{align}
H^{\left(  1\right)  } &  =H_{1}^{\left(  1\right)  }+\Lambda_{1}%
\varphi^{\left(  1|1\right)  },\;\Lambda_{1}=\lambda_{\varphi^{1}}%
-\bar{\lambda}_{1},\nonumber\\
H_{1}^{\left(  1\right)  } &  =H_{1}+\lambda_{\phi^{1}}\phi^{(1|1)}%
,\;H_{1}=H+\bar{\lambda}_{1}\varphi^{\left(  1|1\right)  }\,.\label{2.6}%
\end{align}
The properties
\begin{align}
&  \left\{  \phi^{(1|1)},H^{\left(  1\right)  }+\epsilon\right\}  =\left\{
\phi^{(1|1)},H_{1}^{\left(  1\right)  }+\epsilon\right\}  +O\left(
\Phi^{\left(  1\right)  }\right)  =\left\{  \phi^{(1|1)},H_{1}+\epsilon
\right\}  +O\left(  \Phi^{\left(  1\right)  }\right)  \,,\nonumber\\
&  \left\{  \varphi^{\left(  1|1\right)  },H_{1}+\epsilon\right\}  =O\left(
\Phi^{\left(  1\right)  }\right) \label{2.7}%
\end{align}
are valid.

\subsection{Second stage}

The consistency conditions (\ref{2.1a}) for the primary constraints
$\phi^{(1|1)}$ result in the secondary constraints (the second-stage
constraints)
\[
\left\{  \phi_{\alpha_{1}}^{(1|1)},H_{1}+\epsilon\right\}  \equiv\phi
_{\alpha_{1}}^{(2|1)}=0\,,
\]
which obey the relations
\begin{align}
& \left\{  \varphi^{\left(  1,1\right)  },\phi^{(2|1)}\right\}  =\left\{
\varphi^{\left(  1,1\right)  },\left\{  \phi^{(1|1)},H_{1}+\epsilon\right\}
\right\} \nonumber\\
& \,=\left\{  \left\{  \varphi^{\left(  1,1\right)  },\phi^{(|1)}\right\}
,\,H_{1}+\epsilon\right\}  +\left\{  \phi^{(1|1)},\left\{  \varphi^{\left(
1,1\right)  },H_{1}+\epsilon\right\}  \right\}  =O\left(  \Phi^{\left(
1\right)  },\phi^{(2|1)}\right)  \,.\label{2.9}%
\end{align}
Together with the primary constraints they may form a dependent set of
constraints. We suppose that the matrix $\partial\left(  \Phi^{\left(
1\right)  },\phi^{(2|1)}\right)  /\partial\eta$ has a constant rank,
\begin{equation}
\mathrm{rank}\frac{\partial\left(  \Phi^{\left(  1\right)  },\phi
^{(2|1)}\right)  }{\partial\eta}=m_{1}+m_{2}\leq\left[  \Phi^{\left(
1\right)  }\right]  +\left[  \phi^{(2|1)}\right]  \,.\label{2.8}%
\end{equation}

We now consider the consistency conditions for the constraints $\phi^{(2|1)} $
(one can use the Hamiltonian $H_{1}^{\left(  1\right)  }$ instead of
$H^{\left(  1\right)  }$ in DP)
\begin{align}
& \,\{\phi^{(2|1)},H_{1}^{\left(  1\right)  }+\epsilon\}=\{\phi^{(2|1)}%
,H_{1}+\epsilon\}+C^{(2)}\lambda_{\phi^{1}}=0\,,\nonumber\\
& C^{(2)}=C_{\alpha_{1}\beta_{1}}^{(2)}=\left\{  \phi_{\alpha_{1}}%
^{(2|1)},\phi_{\beta_{1}}^{(1|1)}\right\}  \,.\label{2.10}%
\end{align}
We can see that the matrix $C^{(2)}$ obeys the relation
\begin{align*}
& C_{\alpha_{1}\beta_{1}}^{(2)}=\left\{  \left\{  \phi_{\alpha_{1}}%
^{(1|1)},H_{1}+\epsilon\right\}  ,\phi_{\beta_{1}}^{(1|1)}\right\} \\
& \,=\left\{  \phi_{\alpha_{1}}^{(1|1)},\left\{  H_{1}+\epsilon,\phi
_{\beta_{1}}^{(1|1)}\right\}  \right\}  +\left\{  \left\{  \phi_{\alpha_{1}%
}^{(1|1)},\phi_{\beta_{1}}^{(1|1)}\right\}  ,H_{1}+\epsilon\right\} \\
& \,=\left\{  \phi_{\beta_{1}}^{(2|1)},\phi_{\alpha_{1}}^{(1|1)}\right\}
+O\left(  \Phi^{\left(  1\right)  },\phi^{(2|1)}\right)  =C_{\beta_{1}%
\alpha_{1}}^{(2)}+O\left(  \Phi^{\left(  1\right)  },\phi^{(2|1)}\right)  \,,
\end{align*}
this means that $C^{\left(  2\right)  }$ is symmetric on the constraint
surface $\Phi^{\left(  1\right)  }=\phi^{(2|1)}=0$ (we say that $C^{\left(
2\right)  }$ is $O\left(  \Phi^{\left(  1\right)  },\phi^{(2|1)}\right)
$-symmetric). We suppose that the matrix $C^{\left(  2\right)  }$ has a
constant rank, \textrm{rank\thinspace}$C^{\left(  2\right)  }=$ $r_{2}\,$.
Therefore, there exists a submatrix of size $r_{2}\times r_{2}$, which is
located on the diagonal and is also symmetric, we let denote it by $M_{\mu
_{2}\nu_{2}}^{\left(  2\right)  }\,,\;$ $\left[  \nu_{2}\right]  =\left[
\mu_{2}\right]  =r_{2}\,,$ $\det\,M^{\left(  2\right)  }\neq0.$ We now apply
the $Z-$reorganization to the constraints $\phi^{(1|1)},\phi^{(2|1)}$. The
equation
\[
C^{(2)}Z^{(2)}=O\left(  \Phi^{\left(  1\right)  },\phi^{(2|1)}\right)
\]
has $m_{3}^{\prime\prime}=m_{2}^{\prime}-r_{2}$ linearly independent solutions
$Z_{\sigma_{1}}^{(2)}=Z_{\sigma_{1}}^{(2)\alpha_{1}}$, $[\sigma_{1}%
]=m_{3}^{\prime\prime}\,,$ such that $\det Z_{\sigma_{1}}^{(2)\sigma
_{1}^{\prime}}\neq0$ $(\alpha_{1}=\left(  \mu_{2},\sigma_{1}\right)  ,$
$\left[  \mu_{2}\right]  =r_{2})$. Together with the vectors $Z_{\mu_{2}%
}^{(2)}=Z_{\mu_{2}}^{(2)\alpha_{1}}=\delta_{\mu_{2}}^{\alpha_{1}}$, these
solutions form a set of $m_{2}^{\prime}$ linearly independent vectors. Using a
nonsingular matrix $Z_{2},$ we reorganize the constraints $\phi^{(1|1)}%
,\phi^{(2|1)}$:
\begin{align*}
& \phi^{(1|1)}\overset{Z}{\rightarrow}Z_{2}\phi^{(1|1)}=\left(
\begin{array}
[c]{l}%
\varphi_{\mu_{2}}^{\left(  1|2\right)  }=\phi_{\mu_{2}}^{\left(  1|1\right)
}\\
\Psi_{\sigma_{1}}^{(1|1)}=Z_{\sigma_{1}}^{(2)\alpha_{1}}\phi_{\alpha_{1}%
}^{\left(  1|1\right)  }%
\end{array}
\right)  ,\\
& \phi^{(2|1)}\overset{Z}{\rightarrow}Z_{2}\phi^{(2|1)}=\left(
\begin{array}
[c]{l}%
\varphi_{\mu_{2}}^{\prime\left(  2|2\right)  }=\phi_{\mu_{2}}^{\left(
2|1\right)  }\\
\Psi_{\sigma_{1}}^{\prime(2|1)}=Z_{\sigma_{1}}^{(2)\alpha_{1}}\phi_{\alpha
_{1}}^{\left(  2|1\right)  }%
\end{array}
\right)  ,\\
& Z_{2\alpha_{1}}^{\beta_{1}}=\left(
\begin{array}
[c]{ll}%
\delta_{\mu_{2}}^{\nu_{2}} & 0\\
Z_{\sigma_{1}}^{(2)\nu_{2}} & Z_{\sigma_{1}}^{(2)\sigma_{1}^{\prime}}%
\end{array}
\right)  .
\end{align*}
The new reorganized constraints have the properties
\begin{align}
& \left\{  \varphi^{\prime\left(  2|2\right)  },\varphi^{\left(  1|2\right)
}\right\}  =M^{\left(  2\right)  },\;\;\det M^{\left(  2\right)  }%
\neq0\,,\nonumber\\
& \left\{  \Psi^{\prime(2|1)},\varphi^{\left(  1|2\right)  }\right\}
=O\left(  \Phi^{\left(  1\right)  },\phi^{(2|1)}\right)  ,\;\;\left\{
\Psi^{\prime(2|1)},\Psi^{(1|1)}\right\}  =O\left(  \Phi^{\left(  1\right)
},\phi^{(2|1)}\right)  ,\nonumber\\
& \left\{  \varphi^{\prime\left(  2,2\right)  },\Psi^{(1|1)}\right\}
=O\left(  \Phi^{\left(  1\right)  },\phi^{(2|1)}\right)  \,.\label{2.12}%
\end{align}

In addition, we reorganize the second-stage constraints adding some terms
proportional to the first-stage constraints::
\begin{align}
& \varphi_{\mu_{2}}^{\prime\left(  2|2\right)  }\rightarrow\varphi_{\mu_{2}%
}^{\left(  2|2\right)  }=\varphi_{\mu_{2}}^{\prime\left(  2|2\right)
}-\frac{1}{2}\varphi^{\left(  1|2\right)  }\left[  M^{\left(  2\right)
}\right]  ^{-1}\left\{  \varphi^{\prime\left(  2|2\right)  },\varphi_{\mu_{2}%
}^{\prime\left(  2|2\right)  }\right\}  \,,\nonumber\\
& \Psi_{\sigma_{1}}^{\prime(2|1)}\rightarrow\Psi_{\sigma_{1}}^{(2|1)}%
=\Psi_{\sigma_{1}}^{\prime(2|1)}-\varphi^{\left(  1|2\right)  }\left[
M^{\left(  2\right)  }\right]  ^{-1}\left\{  \varphi^{\prime\left(
2|2\right)  },\Psi_{\sigma_{1}}^{\prime(2|1)}\right\}  \,.\label{2.13}%
\end{align}
We thus have%
\[
\left(
\begin{array}
[c]{l}%
\phi^{(1|1)}\\
\phi^{(2|1)}%
\end{array}
\right)  \overset{Z}{\rightarrow}\left(
\begin{array}
[c]{l}%
\varphi^{\left(  1|2\right)  }\\
\Psi^{(1|1)}\\
\varphi^{\prime\left(  2|2\right)  }\\
\Psi^{\prime(2|1)}%
\end{array}
\right)  \rightarrow\left(
\begin{array}
[c]{l}%
\varphi^{\left(  1|2\right)  }\\
\Psi^{(1|1)}\\
\varphi^{\left(  2|2\right)  }\\
\Psi^{(2|1)}%
\end{array}
\right)  .
\]

The Poisson brackets for the reorganized second-stage constraints
$\varphi^{\left(  2|2\right)  },\Psi^{(2|1)}$ are
\begin{align}
& \left\{  \varphi^{\left(  2|2\right)  },\varphi^{\left(  2|2\right)
}\right\}  =O\left(  \Phi^{\left(  1\right)  },\phi^{(2|1)}\right)
,\;\left\{  \varphi^{\left(  2|2\right)  },\Psi^{(2|1)}\right\}  =O\left(
\Phi^{\left(  1\right)  },\phi^{(2|1)}\right)  ,\nonumber\\
& \left\{  \varphi^{\left(  2,2\right)  },\Psi^{(1|1)}\right\}  =O\left(
\Phi^{\left(  1\right)  },\phi^{(2|1)}\right)  ,\;\left\{  \varphi^{\left(
2|2\right)  },\varphi^{\left(  1|2\right)  }\right\}  =M^{\left(  2\right)
}+O\left(  \Phi^{\left(  1\right)  },\phi^{(2|1)}\right)  ,\nonumber\\
& \left\{  \Psi^{(2|1)},\varphi^{\left(  1|2\right)  }\right\}  =O\left(
\Phi^{\left(  1\right)  },\phi^{(2|1)}\right)  ,\;\left\{  \Psi^{(2|1)}%
,\Psi^{(1|1)}\right\}  =O\left(  \Phi^{\left(  1\right)  },\phi^{(2|1)}%
\right)  \,.\label{2.14}%
\end{align}
At the same time, the reorganized constraints are related to the primary
constraints by
\begin{align*}
\varphi^{\left(  2|2\right)  }  & =\left\{  \varphi^{(1|2)},H_{1}%
+\epsilon\right\}  +O\left(  \Phi^{\left(  1\right)  }\right)  =\left\{
\varphi^{(1|2)},H^{\left(  1\right)  }+\epsilon\right\}  +O\left(
\Phi^{\left(  1\right)  }\right)  \,,\\
\Psi^{(2|1)}  & =\left\{  \Psi^{(1|1)},H_{1}+\epsilon\right\}  +O\left(
\Phi^{\left(  1\right)  }\right)  =\left\{  \Psi^{(1|1)},H^{\left(  1\right)
}+\epsilon\right\}  +O\left(  \Phi^{\left(  1\right)  }\right)  \,.
\end{align*}

One can see that the constraints $\Phi^{\left(  1\right)  },\varphi^{\left(
2|2\right)  }$ are independent.

Taking (\ref{2.8}) into account, we can reorganize the constraints
$\Psi^{(2|1)}$ as follows: $\Psi_{\sigma_{1}}^{(2|1)}\rightarrow\left(
\phi_{\alpha_{2}}^{\left(  2|2\right)  }=\Psi_{\alpha_{2}}^{(2|1)},\chi
_{\rho_{1}}^{\left(  2|1\right)  }=U_{\rho_{1}}^{\left(  2\right)  \sigma_{1}%
}\Psi_{\sigma_{1}}^{(2|1)}\right)  ,\;\sigma_{1}=(\alpha_{2},\rho_{1}),$
$\left[  \phi^{\left(  2|2\right)  }\right]  =m_{2}-r_{2}\equiv m_{3}^{\prime
},$\ where $U_{\rho_{1}}^{\left(  2\right)  }$ are $m_{3}^{\prime\prime}%
-m_{3}^{\prime}\equiv s_{1}$ independent vectors such that the constraints
$\Phi^{\left(  1,2\right)  }$ are independent, and $\chi^{\left(  2|1\right)
}=O\left(  \Phi^{\left(  1\right)  },\varphi^{\left(  2|2\right)  }\right)
$,\ where $\Phi^{\left(  2\right)  }=\left(  \varphi^{\left(  2|2\right)
},\phi^{\left(  2|2\right)  }\right)  ,\;\left[  \Phi^{\left(  2\right)
}\right]  =m_{2}\,;\;$ we then reorganize the constraints $\Psi_{\sigma_{1}%
}^{(1|1)}$: $\Psi_{\sigma_{1}}^{(1|1)}\rightarrow$ $\left(  \phi_{\alpha_{2}%
}^{\left(  1|2\right)  }=\Psi_{\alpha_{2}}^{(1|1)},\chi_{\rho_{1}}^{\left(
1|1\right)  }=U_{\rho_{1}}^{\left(  2\right)  \sigma_{1}}\Psi_{\sigma_{1}%
}^{(1|1)}\right)  ,\;$ $\left[  \chi^{\left(  1|1\right)  }\right]  =s_{1}%
,$\ \ $m_{1}=r_{1}+r_{2}+s_{1}+m_{3}^{\prime},\;m_{2}=r_{2}+m_{3}^{\prime}$ .
The new constraints obey the relations
\begin{align}
& \phi^{\left(  2|2\right)  }=\left\{  \phi^{\left(  1|2\right)  }%
,H_{1}+\epsilon\right\}  +O\left(  \Phi^{\left(  1\right)  }\right)  =\left\{
\phi^{\left(  1|2\right)  },H^{\left(  1\right)  }+\epsilon\right\}  +O\left(
\Phi^{\left(  1\right)  }\right)  ,\nonumber\\
& \left\{  \chi^{\left(  1|1\right)  },H^{\left(  1\right)  }+\epsilon
\right\}  =\chi^{\left(  2|1\right)  }+O\left(  \Phi^{\left(  1\right)
}\right)  =O\left(  \Phi^{\left(  1\right)  },\varphi^{\left(  2|2\right)
}\right)  \,.\label{2.16}%
\end{align}
Thus, the consistency conditions for the\ constraints $\chi^{\left(
1|1\right)  }$\ do not lead to any new constraints. The consistency conditions
for the constraints $\phi^{\left(  2|2\right)  }$ allow us to find the
Lagrange multipliers $\lambda_{\varphi^{2}}$ ,
\begin{equation}
\lambda_{\varphi^{2}}=-\left[  M^{\left(  2\right)  }\right]  ^{-1}\left\{
\phi^{\left(  2|2\right)  },\,H_{1}+\epsilon\right\}  \equiv\bar{\lambda}%
_{2}\,.\label{2.17}%
\end{equation}
It is now useful to represent the Hamiltonian $H^{\left(  1\right)  }$ as
\begin{align}
& H^{\left(  1\right)  }=H_{2}^{\left(  1\right)  }+\lambda_{\chi^{1}}%
\chi^{\left(  1|1\right)  }+\Lambda_{1}\varphi^{\left(  1|1\right)  }%
+\Lambda_{2}\varphi^{\left(  1|2\right)  },\nonumber\\
& H_{2}^{\left(  1\right)  }=H_{2}+\lambda_{\phi^{2}}\phi^{(1|2)}%
,\;\;\Lambda_{2}=\lambda_{\varphi^{2}}-\bar{\lambda}_{2}\,,\nonumber\\
& H_{2}=H_{1}+\bar{\lambda}_{2}\varphi^{\left(  1|2\right)  }=H+\bar{\lambda
}_{1}\varphi^{\left(  1|1\right)  }+\bar{\lambda}_{2}\varphi^{\left(
1|2\right)  }\,.\label{2.18}%
\end{align}

Finally, after the two first stages, we have the following picture:
\[
\begin{array}
[c]{lll}%
\varphi^{\left(  1|1\right)  } & \rightarrow & \bar{\lambda}_{1}\\
\varphi^{\left(  1|2\right)  } & \rightarrow & \varphi^{\left(  2|2\right)
}\rightarrow\bar{\lambda}_{2}\\
\phi^{\left(  1|2\right)  } & \rightarrow & \phi^{\left(  2|2\right)  }\\
\chi^{\left(  1|1\right)  } & \rightarrow & O\left(  \Phi^{\left(  1\right)
},\varphi^{\left(  2|2\right)  }\right)
\end{array}
\,.
\]
The reorganized constraints
\begin{align}
\Phi^{\left(  1\right)  }  & =\left(  \varphi^{\left(  1|1\right)  }%
,\varphi^{\left(  1|2\right)  },\phi^{\left(  1|2\right)  },\chi^{\left(
1|1\right)  }\right)  -\mathrm{primary\;constraints}\,,\nonumber\\
\Phi^{\left(  2\right)  }  & =\left(  \varphi^{\left(  2|2\right)  }%
,\phi^{\left(  2|2\right)  }\right)  -\mathrm{second-stage\;constraints\,}%
\label{2.19}%
\end{align}
are independent. They obey the relations:
\begin{align}
& \left\{  \varphi^{\left(  1|1\right)  },\varphi^{\left(  1|1\right)
}\right\}  =M^{\left(  1\right)  }+O\left(  \Phi^{\left(  1\right)  }\right)
,\nonumber\\
& \left\{  \varphi^{\left(  1|2\right)  },\varphi^{\left(  2|2\right)
}\right\}  =M^{\left(  2\right)  }+O\left(  \Phi^{\left(  1,2\right)
}\right)  ,\nonumber\\
& \left\{  \varphi^{\left(  1|2\right)  },\varphi^{\left(  1|2\right)
}\right\}  =O\left(  \Phi^{\left(  1\right)  }\right)  ,\left\{
\varphi^{\left(  2,2\right)  },\varphi^{\left(  2|2\right)  }\right\}
=O\left(  \Phi^{\left(  1,2\right)  }\right)  ,\nonumber\\
& \left\{  \varphi^{\left(  1|1\right)  },\varphi^{\left(  1|2\right)
}\right\}  =O\left(  \Phi^{\left(  1\right)  }\right)  ,\left\{
\varphi^{\left(  1|1\right)  },\varphi^{\left(  2,2\right)  }\right\}
=O\left(  \Phi^{\left(  1,2\right)  }\right)  ,\nonumber\\
& \left\{  \varphi^{\left(  1|1\right)  },\phi^{\left(  1|2\right)  }\right\}
=O\left(  \Phi^{\left(  1\right)  }\right)  ,\left\{  \varphi^{\left(
1|1\right)  },\chi^{\left(  1|1\right)  }\right\}  =O\left(  \Phi^{\left(
1\right)  }\right)  ,\nonumber\\
& \left\{  \varphi^{\left(  1|2\right)  },\Phi^{\left(  1\right)  }\right\}
=O\left(  \Phi^{\left(  1\right)  }\right)  ,\left\{  \phi^{\left(
1|2\right)  },\Phi^{\left(  1\right)  }\right\}  =O\left(  \Phi^{\left(
1\right)  }\right)  ,\nonumber\\
& \left\{  \chi^{\left(  1|1\right)  },\Phi^{\left(  1\right)  }\right\}
=O\left(  \Phi^{\left(  1\right)  }\right)  ,\nonumber\\
& \left\{  \varphi^{\left(  2,2\right)  },\phi^{\left(  1|2\right)  }\right\}
=O\left(  \Phi^{\left(  1,2\right)  }\right)  ,\left\{  \varphi^{\left(
2,2\right)  },\chi^{\left(  1|1\right)  }\right\}  =O\left(  \Phi^{\left(
1,2\right)  }\right)  ,\nonumber\\
& \left\{  \varphi^{\left(  2,2\right)  },\phi^{\left(  2|2\right)  }\right\}
=O\left(  \Phi^{\left(  1,2\right)  }\right)  ,\left\{  \phi^{\left(
2,2\right)  },\Phi^{\left(  1\right)  }\right\}  =O\left(  \Phi^{\left(
1,2\right)  }\right)  ,\label{2.20}%
\end{align}
and
\begin{align*}
& \left\{  \varphi^{\left(  2|2\right)  },H_{2}+\varepsilon\right\}  =O\left(
\Phi^{\left(  1,2\right)  }\right)  ,\\
& \left\{  \phi^{\left(  2|2\right)  },H^{\left(  1\right)  }+\varepsilon
\right\}  =\left\{  \phi^{\left(  2|2\right)  },H_{2}+\varepsilon\right\}
+O\left(  \Phi^{\left(  1,2\right)  }\right)  ,\\
& \left\{  \varphi^{\left(  1|1\right)  },H_{i}+\varepsilon\right\}  =O\left(
\Phi^{\left(  1\right)  }\right)  ,\\
& \left\{  \chi^{\left(  1|1\right)  },H_{i}+\varepsilon\right\}  =O\left(
\Phi^{\left(  1\right)  },\varphi^{\left(  2|2\right)  }\right)  ,\\
& \left\{  \varphi^{\left(  1|2\right)  },H_{i}+\varepsilon\right\}
=\varphi^{\left(  2,2\right)  }+O\left(  \Phi^{\left(  1\right)  }\right)  ,\\
& \left\{  \phi^{\left(  1|2\right)  },H_{i}+\varepsilon\right\}
=\phi^{\left(  2,2\right)  }+O\left(  \Phi^{\left(  1\right)  }\right)
,\;i=1,2,
\end{align*}
where the Hamiltonians $H_{i}\,,$ are given by Eqs. (\ref{2.6}) and
(\ref{2.18}). The commutation relations $\left\{  \phi^{\left(  2,2\right)
},\phi^{\left(  2|2\right)  }\right\}  $ remain unknown at this stage.

It follows from (\ref{2.20}) that $\varphi^{\left(  1|2\right)  }%
,\varphi^{\left(  2|2\right)  }$ are SCC.

Considering the consistency conditions for the constraints $\phi^{\left(
2|2\right)  },$ we can use the Hamiltonian $H_{2}$ instead of $H^{\left(
1\right)  }$. These consistency conditions result in the third-stage
constraints$,$%
\begin{equation}
\phi^{\left(  3|2\right)  }\equiv\left\{  \phi^{\left(  2|2\right)  }%
,H_{2}+\varepsilon\right\}  =0.\label{2.21}%
\end{equation}
The functions $\phi^{\left(  3|2\right)  }$ must be analyzed similarly to the
previous consideration.

\subsection{$p$-th stage (induction hypothesis)}

We are now going to prove that the above-refined DP formulated for two stages
can be continued producing \ similar structures for any-stage constraints. The
proof is by induction. The induction hypothesis is formulated as follows.

Suppose that after any $l\leq p$ stages of the refined DP, the constraints
$\Phi^{\left(  1,...,l\right)  }$ and the total Hamiltonian $H^{\left(
1\right)  }$ can be reorganized as%

\begin{align}
& \Phi^{\left(  i\right)  }=\left(  \varphi_{\mu_{u}}^{\left(  i|u\right)
},\,\phi_{\alpha_{l}}^{\left(  i|l\right)  },\,\chi_{\rho_{a}}^{\left(
i|a\right)  }\right)  ;\;\left[  \Phi^{\left(  i\right)  }\right]  =m_{i}%
=\sum_{j=i}^{l}r_{j}+\sum_{a=i}^{l-1}s_{a}+m_{l+1}^{\prime}\,,\nonumber\\
& 1\leq i\leq l,\;i\leq u\leq l,\;i\leq a\leq l-1\,,\nonumber\\
& \left[  \varphi^{\left(  i|u\right)  }\right]  =r_{u}\,,\;\left[
\chi^{\left(  i|a\right)  }\right]  =s_{a}\,,\;\left[  \phi^{\left(
i|l\right)  }\right]  =m_{l+1}^{\prime}\,,\nonumber\\
& H^{\left(  1\right)  }=H_{l}^{\left(  1\right)  }+\sum_{u=1}^{l}\Lambda
_{u}\varphi^{\left(  1|u\right)  }+\sum_{a=1}^{l-1}\lambda_{\chi^{a}}%
\chi^{\left(  1|a\right)  },\;\Lambda_{u}=\lambda_{\varphi^{u}}-\bar{\lambda
}_{u}\,,\nonumber\\
& H_{l}^{\left(  1\right)  }=H_{l}+\lambda_{\phi^{l}}\phi^{\left(  1|l\right)
},\;H_{l}=H_{l-1}+\bar{\lambda}_{l}\varphi^{\left(  1|l\right)  }=H+\sum
_{u=1}^{l}\bar{\lambda}_{u}\varphi^{\left(  1|u\right)  }\,,\nonumber\\
& \bar{\lambda}_{u}=-\left[  M^{\left(  u\right)  }\right]  ^{-1}\left\{
\varphi^{\left(  u|u\right)  },H_{u-1}+\varepsilon\right\}  \,,\;M^{\left(
u\right)  }=\left\{  \varphi^{\left(  u|u\right)  },\varphi^{\left(
1|u\right)  }\right\}  ,\nonumber\\
& \det M^{\left(  u\right)  }\neq0\,,\;\;\left(  H_{0}\equiv H\right)
\,.\label{2.22}%
\end{align}
All the constraints $\Phi^{\left(  1,...,l\right)  }$ are independent. In
passing from $l$-th stage ($l\leq p-1)$ to $\left(  l+1\right)  $-th stage,
the only constraints to be reorganized are $\phi^{\left(  i|l\right)  }$.

The constraints obey the relations
\begin{align}
& u<v:\;\left\{  \varphi^{\left(  i|u\right)  },\varphi^{\left(  j|v\right)
}\right\}  =\left\{
\begin{array}
[c]{l}%
O\left(  \Phi^{\left(  1,...,i+j-1\right)  }\right)  ,\;i+j\leq u+1\\
O\left(  \Phi^{\left(  1,...,u\right)  }\right)  ,\;i+j>u+1,\;j\leq u\\
O\left(  \Phi^{\left(  1,...,j\right)  }\right)  ,\;j>u\,,
\end{array}
\right. \nonumber\\
& \left\{  \varphi^{\left(  i|u\right)  },\varphi^{\left(  j|u\right)
}\right\}  =\left\{
\begin{array}
[c]{l}%
O\left(  \Phi^{\left(  1,...,i+j-1\right)  }\right)  ,\;i+j\leq u+1\\
\left(  -1\right)  ^{u-i}M^{\left(  u\right)  }+O\left(  \Phi^{\left(
1,...,u\right)  }\right)  ,\;i+j=u+1\\
O\left(  \Phi^{\left(  1,...,j\right)  }\right)  ,\;i+j>u+1\,,
\end{array}
\right.  \,\label{2.26}%
\end{align}
i.e., $\varphi^{\left(  i|u\right)  }$ are\ of the second class,
\begin{align}
& \left\{  T_{l}^{\left(  i\right)  },T_{l}^{\left(  j\right)  }\right\}
=O\left(  \Phi^{\left(  1,...,i+j-1\right)  }\right)  ,\;i+j\leq
l+1\,,\nonumber\\
& T_{l}^{\left(  i\right)  }\equiv\left(  \phi^{\left(  i|l\right)  }%
,\;\chi^{\left(  i|a\right)  },\;i\leq a\leq l-1\right)  \,;\nonumber\\
& \left\{  \varphi^{\left(  i|u\right)  },T_{l}^{\left(  j\right)  }\right\}
=\left\{
\begin{array}
[c]{l}%
O\left(  \Phi^{\left(  1,...,i+j-1\right)  }\right)  ,\;i+j\leq u+1\\
O\left(  \Phi^{\left(  1,...,u\right)  }\right)  ,\;i+j>u+1,\;j\leq u\\
O\left(  \Phi^{\left(  1,...,j\right)  }\right)  ,\;j>u,
\end{array}
\right. \label{2.26a}%
\end{align}
and
\begin{align*}
& \left\{  \varphi^{\left(  i|u\right)  },H_{j}+\varepsilon\right\}
=\varphi^{\left(  i+1|u\right)  }+O\left(  \Phi^{\left(  1,...,i\right)
}\right)  ,\;j\geq i,\;u>i\,,\\
& \left\{  \varphi^{\left(  i|i\right)  },H_{j}+\varepsilon\right\}  =O\left(
\Phi^{\left(  1,...,i\right)  }\right)  ,\;j\geq i\,,\\
& \left\{  \phi^{\left(  i|l\right)  },H_{j}+\varepsilon\right\}
=\phi^{\left(  i+1|l\right)  }+O\left(  \Phi^{\left(  1,...,i\right)
}\right)  ,\;i\leq j,\;i\leq l-1\,,\\
& \left\{  \chi^{\left(  i|a\right)  },H_{j}+\varepsilon\right\}
=\chi^{\left(  i+1|a\right)  }+O\left(  \Phi^{\left(  1,...,i\right)
}\right)  ,\;i\leq j,\;i\leq l-1,\;a\geq i+1\,,\\
& \left\{  \chi^{\left(  i|i\right)  },H_{j}+\varepsilon\right\}  =O\left(
\Phi^{\left(  1,...,i\right)  },\varphi^{\left(  i+1|i+1\right)  }\right)
,\;i\leq j,\;i\leq l-1\,,\\
& \left\{  \phi^{\left(  l|l\right)  },H^{\left(  1\right)  }+\varepsilon
\right\}  =\left\{  \phi^{\left(  l|l\right)  },H_{l}+\varepsilon\right\}
+O\left(  \Phi^{\left(  1,...,l\right)  }\right)  \,.
\end{align*}
The second-stage constraints satisfy this hypothesis.

\subsection{$\left(  p+1\right)  $-th stage}

Let us consider the $\left(  p+1\right)  $-th stage of the refined DP. The
consistency conditions for the constraints $\phi^{\left(  p|p\right)  }$
result in the $\left(  p+1\right)  $-th stage constraints,
\begin{equation}
\left\{  \phi^{\left(  p|p\right)  },H_{p}+\varepsilon\right\}  \equiv
\phi^{\prime\left(  p+1|p\right)  }=0\,.\label{2.27}%
\end{equation}
The $\left(  p+1\right)  $-th stage constraints $\phi^{\prime\left(
p+1|p\right)  }$ $=\phi_{\alpha_{p}}^{\prime\left(  p+1|p\right)  }$ together
with constraints of the previous stages may form a dependent set of
constraints. We suppose that the matrix $\partial\!\left(  \Phi^{\left(
1,...,p\right)  },\phi^{\prime(p+1|p)}\right)  \!/\partial\eta$ has a constant
rank,
\begin{equation}
\mathrm{rank}\frac{\partial\left(  \Phi^{\left(  1,...,p\right)  }%
,\phi^{\prime(p+1|p)}\right)  }{\partial\eta}=\sum_{i=1}^{p}m_{i}+m_{p+1}%
\leq\left[  \Phi^{\left(  1,...,p\right)  }\right]  +\left[  \phi
^{\prime(p+1|p)}\right]  .\label{2.28}%
\end{equation}
We first reorganize the constraints $\phi^{\prime(p+1|p)}$ to $\phi^{\left(
p+1|p\right)  }$\ as follows:
\begin{align*}
& \phi_{\alpha_{p}}^{\prime\left(  p+1|p\right)  }\rightarrow\phi_{\alpha_{p}%
}^{\left(  p+1|p\right)  }=\phi_{\alpha_{p}}^{\prime\left(  p+1|p\right)  }\\
& -\sum_{u=2}^{p}\sum_{i=1}^{u-1}\varphi^{\left(  i|u\right)  }\left[
M^{\left(  u\right)  }\right]  ^{-1}\left\{  \varphi^{\left(  u+1-i|u\right)
},\phi_{\alpha_{p}}^{\prime\left(  p+1|p\right)  }\right\}  .
\end{align*}
The new constraints $\phi^{\left(  p+1|p\right)  }$ obey the relations
\begin{align}
& \phi^{\left(  p+1|p\right)  }=\left\{  \phi^{\left(  p|p\right)  }%
,H_{p}+\varepsilon\right\}  +O\left(  \Phi^{\left(  1,...,p\right)  }\right)
\,,\nonumber\\
& \left\{  \varphi^{\left(  i|u\right)  },\phi^{\left(  p+1|p\right)
}\right\}  =O\left(  \Phi^{\left(  1,...,p\right)  }\right)  ,\;2\leq i\leq
p,\;i\leq u\leq p\,,\nonumber\\
& \left\{  \varphi^{\left(  1|u\right)  },\phi^{\left(  p+1|p\right)
}\right\}  =\left\{  \varphi^{\left(  1|u\right)  },\left\{  \phi^{\left(
p|p\right)  },H_{p}+\varepsilon\right\}  \right\}  +O\left(  \Phi^{\left(
1,...,p\right)  }\right)  \nonumber\\
& \,=\left\{  \left\{  \varphi^{\left(  1|u\right)  },\phi^{\left(
p|p\right)  }\right\}  ,H_{p}+\varepsilon\right\}  +\left\{  \phi^{\left(
p|p\right)  },\left\{  \varphi^{\left(  1|u\right)  },H_{p}+\varepsilon
\right\}  \right\}  \nonumber\\
& +O\left(  \Phi^{\left(  1,...,p\right)  }\right)  =O\left(  \Phi^{\left(
1,...,p\right)  },\phi^{\left(  p+1|p\right)  }\right)
,\;u=1,...,p,\nonumber\\
& \left\{  \chi^{\left(  1|a\right)  },\phi^{\left(  p+1|p\right)  }\right\}
=O\left(  \Phi^{\left(  1,...,p\right)  },\phi^{\left(  p+1|p\right)
}\right)  ,\;a=1,...,p-1\,,\nonumber\\
& \left\{  \phi^{\left(  p+1|p\right)  },H^{\left(  1\right)  }+\varepsilon
\right\}  =\left\{  \phi^{\left(  p+1|p\right)  },H_{p}^{\left(  1\right)
}+\varepsilon\right\}  \nonumber\\
& \,+O\left(  \Phi^{\left(  1,...,p\right)  },\phi^{\left(  p+1|p\right)
}\right)  .\label{2.29}%
\end{align}

Consider the consistency conditions for $\phi^{\left(  p+1|p\right)  }$ (we
can use $H_{p}^{\left(  1\right)  }$ instead of $H^{\left(  1\right)  }$ in
DP)
\[
\{\phi^{(p+1|p)},H_{p}+\epsilon\}+C^{(p+1)}\lambda_{p}=0,\;C^{(p+1)}%
=C_{\alpha_{p}\beta_{p}}^{(p+1)}=\left\{  \phi_{\alpha_{p}}^{(p+1|p)}%
,\phi_{\beta_{p}}^{(1|p)}\right\}  .
\]
We can see that the matrix $C^{(p+1)}$ and all the matrices $\left\{
\phi_{\alpha_{p}}^{(p+2-i|p)},\phi_{\beta_{p}}^{(i|p)}\right\}  ,\;i=2,...,p$
coincide up to a sign and are (anti)symmetric on the constraint surface
$\Phi^{\left(  1,...,p\right)  }=\phi^{\left(  p+1|p\right)  }=0$:
\begin{align*}
& C_{\alpha_{p}\beta_{p}}^{(p+1)}=\left\{  \left\{  \phi_{\alpha_{p}}%
^{(p|p)},H_{p}+\epsilon\right\}  ,\phi_{\beta_{p}}^{(1|p)}\right\}  +O\left(
\Phi^{\left(  1,...,p\right)  }\right)  =-\left\{  \phi_{\alpha_{p}}%
^{(p|p)},\phi_{\beta_{p}}^{(2|p)}\right\} \\
& \,+O\left(  \Phi^{\left(  1,...,p\right)  },\phi^{\left(  p+1|p\right)
}\right)  =\cdots=\left(  -1\right)  ^{i}\left\{  \phi_{\alpha_{p}%
}^{(p+1-i|p)},\phi_{\beta_{p}}^{(i+1|p)}\right\} \\
& \,+O\left(  \Phi^{\left(  1,...,p\right)  },\phi^{\left(  p+1|p\right)
}\right)  =\cdots=\left(  -1\right)  ^{p}\left\{  \phi_{\alpha_{p}}%
^{(1|p)},\phi_{\beta_{p}}^{(p+1|p)}\right\} \\
& \,+O\left(  \Phi^{\left(  1,...,p\right)  },\phi^{\left(  p+1|p\right)
}\right)  =\left(  -1\right)  ^{p+1}C_{\beta_{p}\alpha_{p}}^{(p+1)}+O\left(
\Phi^{\left(  1,...,p\right)  },\phi^{\left(  p+1|p\right)  }\right)  \,.
\end{align*}
Here, we have used the Jacobi identity and Eqs. (\ref{2.22}, \ref{2.26},
\ref{2.26a}, \ref{2.29}).

We suppose that the matrix $C^{\left(  p+1\right)  }$ has a constant rank,
\textrm{rank}$\,C^{\left(  p+1\right)  }=r_{p+1}$. We then perform the
$Z-$reorganization. Namely, we consider the equation
\[
C^{(p+1)}Z^{(p+1)}=O\left(  \Phi^{\left(  1,...,p\right)  },\phi^{\left(
p+1|p\right)  }\right)  ,\;
\]
which has $m_{p+2}^{\prime\prime}=m_{p+1}^{\prime}-r_{p+1}$ linearly
independent solutions $Z_{\sigma_{p}}^{(p+1)}$=$Z_{\sigma_{p}}^{(p+1)\alpha
_{p}}$, $[\sigma_{p}]=m_{p+2}^{\prime\prime}$, such that $\det Z_{\sigma_{p}%
}^{(p+1)\sigma_{p}^{\prime}}\neq0$ $(\alpha_{p}=\left(  \mu_{p+1},\sigma
_{p}\right)  ,$ $\left[  \mu_{p+1}\right]  =r_{p+1})$. These solutions
together with the vectors $Z_{\mu_{p+1}}^{(p+1)}=Z_{\mu_{p+1}}^{(p+1)\alpha
_{p}}=\delta_{\mu_{p+1}}^{\alpha_{p}}$ form the set of $m_{p+1}^{\prime}$
linearly independent vectors. We reorganize the constraints $\phi
^{(i|p)},i=1,...,p+1 $ using a nonsingular matrix $Z_{p+1}$:
\begin{align*}
& \phi^{(i|p)}\rightarrow Z_{p+1}\phi^{(i|p)}=\left(
\begin{array}
[c]{l}%
\varphi_{\mu_{p+1}}^{\prime\left(  i|p+1\right)  }=\phi_{\mu_{p+1}}^{\left(
i|p\right)  }\\
\Psi_{\sigma_{p}}^{\prime(i|p)}=Z_{\sigma_{p}}^{(p+1)\alpha_{p}}\phi
_{\alpha_{p}}^{\left(  i|p\right)  }%
\end{array}
\right)  ,\;\alpha_{p}=\left(  \mu_{p+1},\sigma_{p}\right)  \,,\\
& Z_{p+1}=Z_{p+1\alpha_{p}}^{\beta_{p}}=\left(
\begin{array}
[c]{ll}%
\delta_{\mu_{p+1}}^{\nu_{p+1}} & 0\\
Z_{\sigma_{p}}^{(p+1)\nu_{p+1}} & Z_{\sigma_{p}}^{(p+1)\sigma_{p}^{\prime}}%
\end{array}
\right)  ,\;\beta_{p}=\left(  \nu_{p+1},\sigma_{p}^{\prime}\right)  \,.
\end{align*}

Next, a set of additional reorganizations must be carried out (by adding some
previous-stage constraints). We first reorganize the constraints
$\varphi^{\prime\left(  i|p+1\right)  }$:
\begin{align*}
& \varphi_{\mu_{p+1}}^{\prime\left(  1|p+1\right)  }\rightarrow\varphi
_{\mu_{p+1}}^{\prime\prime\left(  1|p+1\right)  }=\varphi_{\mu_{p+1}}%
^{\prime\left(  1|p+1\right)  }\,,\\
& \varphi_{\mu_{p+1}}^{\prime\left(  i|p+1\right)  }\rightarrow\varphi
_{\mu_{p+1}}^{\prime\prime\left(  i|p+1\right)  }=\varphi_{\mu_{p+1}}%
^{\prime\left(  i|p+1\right)  }-\varphi^{\prime\left(  1|p+1\right)  }\left(
\left\{  \varphi^{\prime\left(  p+1|p+1\right)  },\varphi^{\prime\left(
1|p+1\right)  }\right\}  \right)  ^{-1}\\
& \,\times\left\{  \varphi^{\prime\left(  p+1|p+1\right)  },\varphi_{\mu
_{p+1}}^{\prime\left(  i|p+1\right)  }\right\}  ,\;i=2,...,p\,;\\
& \varphi_{\mu_{p+1}}^{\prime\left(  p+1|p+1\right)  }\rightarrow\varphi
_{\mu_{p+1}}^{\prime\prime\left(  p+1|p+1\right)  }=\varphi_{\mu_{p+1}%
}^{\prime\left(  p+1|p+1\right)  }\\
& -\frac{1}{2}\varphi^{\prime\left(  1|p+1\right)  }\left(  \left\{
\varphi^{\prime\left(  p+1|p+1\right)  },\varphi^{\prime\left(  1|p+1\right)
}\right\}  \right)  ^{-1}\left\{  \varphi^{\prime\left(  p+1|p+1\right)
},\varphi_{\mu_{p+1}}^{\prime\left(  p+1|p+1\right)  }\right\}  ;
\end{align*}
then, we reorganize the constraints $\varphi^{\prime\prime\left(
i|p+1\right)  }$:
\begin{align*}
& \varphi_{\mu_{p+1}}^{\prime\prime\left(  i|p+1\right)  }\rightarrow
\varphi_{\mu_{p+1}}^{\prime\prime\prime\left(  i|p+1\right)  }=\varphi
_{\mu_{p+1}}^{\prime\prime\left(  i|p+1\right)  },\;i=1,2,p+1\,,\\
& \varphi_{\mu_{p+1}}^{\prime\prime\left(  i|p+1\right)  }\rightarrow
\varphi_{\mu_{p+1}}^{\prime\prime\prime\left(  i|p+1\right)  }=\varphi
_{\mu_{p+1}}^{\prime\prime\left(  i|p+1\right)  }-\varphi^{\prime\prime\left(
2|p+1\right)  }\left(  \left\{  \varphi^{\prime\prime\left(  p|p+1\right)
},\varphi^{\prime\prime\left(  2|p+1\right)  }\right\}  \right)  ^{-1}\\
& \,\times\left\{  \varphi^{\prime\prime\left(  p|p+1\right)  },\varphi
_{\mu_{p+1}}^{\prime\prime\left(  i|p+1\right)  }\right\}  ,\;i=3,...,p-1\,;\\
& \varphi_{\mu_{p+1}}^{\prime\prime\left(  p|p+1\right)  }\rightarrow
\varphi_{\mu_{p+1}}^{\prime\prime\prime\left(  p|p+1\right)  }=\varphi
_{\mu_{p+1}}^{\prime\prime\left(  p|p+1\right)  }\\
& -\frac{1}{2}\varphi^{\prime\prime\left(  2|p+1\right)  }\left(  \left\{
\varphi^{\prime\prime\left(  p|p+1\right)  },\varphi^{\prime\prime\left(
2|p+1\right)  }\right\}  \right)  ^{-1}\left\{  \varphi^{\prime\prime\left(
p|p+1\right)  },\varphi_{\mu_{p+1}}^{\prime\prime\left(  p|p+1\right)
}\right\}  ;
\end{align*}
etc.. This set of reorganizations ends up with\ producing the $\varphi$'s with
$d$ primes, where $d$ is equal to the integer part of $\left(  p+1\right)  /2$
$\;(d=\left[  \left(  p+1\right)  /2\right]  )$,
\begin{align*}
& \varphi_{\mu_{p+1}}^{\prime\prime...\left(  i|p+1\right)  }\rightarrow
\varphi_{\mu_{p+1}}^{\prime\prime...\prime\left(  i|p+1\right)  }=\varphi
_{\mu_{p+1}}^{\prime\prime...\left(  i|p+1\right)  }%
,\;i=1,...,d,p+3-d,...,p+1;\\
& \varphi_{\mu_{p+1}}^{\prime\prime...\left(  i|p+1\right)  }\rightarrow
\varphi_{\mu_{p+1}}^{\prime\prime...\prime\left(  i|p+1\right)  }=\varphi
_{\mu_{p+1}}^{\prime\prime...\left(  i|p+1\right)  }-\varphi^{\prime
\prime...\left(  l|p+1\right)  }\\
& \times\left(  \left\{  \varphi^{\prime\prime...\left(  p+2-d|p+1\right)
},\varphi^{\prime\prime...\left(  d|p+1\right)  }\right\}  \right)
^{-1}\left\{  \varphi^{\prime\prime...\left(  p+2-d|p+1\right)  },\varphi
_{\mu_{p+1}}^{\prime\prime...\left(  i|p+1\right)  }\right\}  ,\\
& i=d+1,p+1-d;\\
& \varphi_{\mu_{p+1}}^{\prime\prime...\left(  p+2-d|p+1\right)  }%
\rightarrow\varphi_{\mu_{p+1}}^{\prime\prime...\prime\left(  p+2-d|p+1\right)
}=\varphi_{\mu_{p+1}}^{\prime\prime...\left(  p+2-d|p+1\right)  }-\frac{1}%
{2}\varphi^{\prime\prime...\left(  d|p+1\right)  }\\
& \times\left(  \left\{  \varphi^{\prime\prime...\left(  p+2-d|p+1\right)
},\varphi^{\prime\prime...\left(  d|p+1\right)  }\right\}  \right)
^{-1}\left\{  \varphi^{\prime\prime...\left(  p|p+1\right)  },\varphi
_{\mu_{p+1}}^{\prime\prime...\left(  p+2-d|p+1\right)  }\right\}  \,.
\end{align*}
In what follows, we omit all the primes such that $\varphi^{\left(
i|p+1\right)  }$ are the final reorganized constraints. These constraints
satisfy the relations (\ref{2.22})-(\ref{2.26a}) with $p\rightarrow p+1$.

We now reorganize the constraints $\Psi^{\prime(i|p)}$:
\begin{align*}
& \Psi_{\sigma_{p}}^{\prime(1|p)}\rightarrow\Psi_{\sigma_{p}}^{(1|p)}%
=\Psi_{\sigma_{p}}^{\prime(1|p)};\;\;\Psi_{\sigma_{p}}^{\prime(i|p)}%
\rightarrow\Psi_{\sigma_{p}}^{(i|p)}=\Psi_{\sigma_{p}}^{\prime(i|p)}%
-\sum_{j=1}^{i-1}\varphi^{\left(  j|p+1\right)  }\\
& \,\times\left(  \left\{  \varphi^{\left(  p+2-j|p+1\right)  },\varphi
^{\left(  j|p+1\right)  }\right\}  \right)  ^{-1}\left\{  \varphi^{\left(
p+2-j|p+1\right)  },\Psi_{\sigma_{p}}^{\prime(i|p)}\right\}
\,,\;i=2,...,p+1\,.
\end{align*}
The constraints $\Psi^{(i|p)}$ have the properties
\begin{align*}
& \left\{  \varphi^{\left(  i|u\right)  },\Psi^{(j|p)}\right\}  =\left\{
\begin{array}
[c]{l}%
O\left(  \Phi^{\left(  1,...,i+j-1\right)  }\right)  ,\;i+j\leq u+1\\
O\left(  \Phi^{\left(  1,...,u\right)  }\right)  ,\;i+j>u+1,\;j\leq u\\
O\left(  \Phi^{\left(  1,...,j\right)  }\right)  ,\;j>u,
\end{array}
\right. \\
& \left\{  \Psi^{(i|p)},\Psi^{(j|p)}\right\}  =O\left(  \Phi^{\left(
1,...,i+j-1\right)  }\right)  ,\;i+j\leq\left(  p+1\right)  +1\,.
\end{align*}

We can see that the constraints $\left(  \Phi^{\left(  1,...,p\right)
},\varphi^{\left(  p+1|p+1\right)  }\right)  $ are independent. Taking this
and Eq. (\ref{2.28}) into account, we reorganize the constraints
$\Psi^{(p+1|p)}$: $\Psi_{\sigma_{p}}^{(p+1|p)}\rightarrow\left(  \phi
_{\alpha_{p+1}}^{\left(  p+1|p+1\right)  }=\Psi_{\alpha_{p+1}}^{(p+1|p+1)}%
,\,\chi_{\rho_{p}}^{\left(  p+1|p\right)  }=U_{\rho_{p}}^{\left(  p+1\right)
\sigma_{p}}\Psi_{\sigma_{p}}^{(p+1|p)}\right)  ,\;\sigma_{p}=(\alpha
_{p+1},\rho_{p}),$ $\left[  \phi^{\left(  p+1|p+1\right)  }\right]
=m_{p+1}-r_{p+1}\equiv m_{p+2}^{\prime},$ $U_{\rho_{p}}^{\left(  p+1\right)
}$ is a set of independent vectors, $\left[  U^{\left(  p+1\right)  }\right]
=\left[  \chi^{\left(  p+1|p\right)  }\right]  =m_{p+2}^{\prime\prime}%
-m_{p+2}^{\prime}$ $\equiv s_{p},\,$ such that the constraints $\Phi^{\left(
1,...,p+1\right)  }$, where $\Phi^{\left(  p+1\right)  }=\left(
\varphi^{\left(  p+1|p+1\right)  },\phi^{\left(  p+1|p+1\right)  }\right)
$,$\;\left[  \Phi^{\left(  p+1\right)  }\right]  =m_{p+1}$, are independent
and $\chi^{\left(  p+1|p\right)  }=O\left(  \Phi^{\left(  1,...,p\right)
},\varphi^{\left(  p+1|1,...,p+1\right)  }\right)  $.$\;$Then, we reorganize
the constraints $\Psi^{(i|p)},\;i=1,...,p$:
\begin{align*}
& \Psi_{\sigma_{p}}^{(i|p)}\rightarrow\left(  \phi_{\alpha_{p+1}}^{\left(
i|p+1\right)  }=\Psi_{\alpha_{p+1}}^{(i|p+1)},\chi_{\rho_{p}}^{\left(
i|p\right)  }=U_{\rho_{p}}^{\left(  p+1\right)  \sigma_{p}}\Psi_{\sigma_{p}%
}^{(i|p)}\right)  \,,\\
& \left[  \chi^{\left(  i|p\right)  }\right]  =s_{p}\,,\;m_{i}=\sum
_{j=i}^{p+1}r_{j}+\sum_{a=i}^{p}s_{a}+m_{p+2}^{\prime}\,.
\end{align*}

The consistency conditions for the\ constraints $\chi^{\left(  p+1|p\right)  }
$\ do not produce any new constraints.

Introducing the Hamiltonian $H_{p+1}^{\left(  1\right)  }\,,$%
\begin{align*}
& H^{\left(  1\right)  }=H_{p+1}^{\left(  1\right)  }+\sum_{u=1}^{p+1}%
\Lambda_{u}\varphi^{\left(  1|u\right)  }+\sum_{a=1}^{p}\lambda_{\chi^{a}}%
\chi^{\left(  1|a\right)  },\;\Lambda_{u}=\lambda_{\varphi^{u}}-\bar{\lambda
}_{u}\,,\\
& H_{p+1}^{\left(  1\right)  }=H_{p+1}+\lambda_{\phi^{p+1}}\phi^{\left(
1|p+1\right)  },\;H_{p+1}=H_{p}+\bar{\lambda}_{p+1}\varphi^{\left(
1|p+1\right)  }\,,\\
& \bar{\lambda}_{p+1}=-\left[  M^{\left(  p+1\right)  }\right]  ^{-1}\left\{
\varphi^{\left(  p+1|p+1\right)  },H_{p}+\varepsilon\right\}  \,,\\
& M^{\left(  p+1\right)  }=M_{\mu_{p+1}\nu_{p+1}}^{\left(  p+1\right)
}=C_{\mu_{p+1}\nu_{p+1}}^{\left(  p+1\right)  }\,,
\end{align*}
we can straightforwardly verify that all the constraints reorganized up to the
$\left(  p+1\right)  $-th stage satisfy the induction hypothesis.

\subsection{Final stage}

DP ends up at the $k$-th stage if $\left[  \phi^{\left(  k|k\right)  }\right]
=m_{k+1}^{\prime}=0$. The latter means that $\left[  \phi^{\left(  i|k\right)
}\right]  =0$,$\;$ $i=1,...,k-1,$ and $T_{k}^{\left(  i\right)  }=\left(
\chi^{\left(  i|a\right)  }\right)  ,$ see (\ref{2.26}) . The constraints
$\chi^{\left(  i|a\right)  }$ therefore commute as follows :
\[
\left\{  \chi^{\left(  i|a\right)  },\chi^{\left(  j|b\right)  }\right\}
=O\left(  \Phi^{\left(  1,...,i+j-1\right)  }\right)  ,\;i+j\leq k+1\,.
\]
Considering the relations
\begin{align*}
& O\left(  \Phi^{\left(  1,...,k\right)  }\right)  =\left\{  \chi^{\left(
1|a\right)  },\left\{  \chi^{\left(  k-1|k-1\right)  },H_{k}+\epsilon\right\}
\right\} \\
& \,=\left\{  \chi^{\left(  k-1|k-1\right)  },\chi^{\left(  2|a\right)
}\right\}  +O\left(  \Phi^{\left(  1,...,k\right)  }\right)  \,,
\end{align*}
we obtain
\[
\left\{  \chi^{\left(  i|a\right)  },\chi^{\left(  j|b\right)  }\right\}
=O\left(  \Phi^{\left(  1,...,k\right)  }\right)  ,\;i+j=k+2\,.
\]
Then, considering the double commutator $\left\{  \chi^{\left(  2|a\right)
},\left\{  \chi^{\left(  k-1|k-1\right)  },H_{k}+\epsilon\right\}  \right\}
,$ we obtain
\[
\left\{  \chi^{\left(  i|a\right)  },\chi^{\left(  j|b\right)  }\right\}
=O\left(  \Phi^{\left(  1,...,k\right)  }\right)  ,\;i+j=k+3\,,
\]
and so on. We finally have
\begin{equation}
\left\{  \chi^{\left(  i|a\right)  },\chi^{\left(  j|b\right)  }\right\}
=\left\{
\begin{array}
[c]{l}%
O\left(  \Phi^{\left(  1,...,i+j-1\right)  }\right)  ,\;i+j<k+1\\
O\left(  \Phi^{\left(  1,...,k\right)  }\right)  ,\;i+j\geq k+1
\end{array}
\right.  .\label{2.31}%
\end{equation}
In particular, we can conclude (only at the last stage!)\ that\ all the
constraints $\chi^{\left(  i|a\right)  }$ are FCC.

\section{Summary}

We here summarize the constraint structure consistent with DP. To make things
more clear, we repeat some points.

It is possible to reorganize the complete set of constraints obtained\ by DP
to the following form:\
\begin{align}
& \Phi=\Phi^{\left(  1,...,k\right)  }=\left(  \Phi^{\left(  i\right)
}\right)  \,,\;\ \Phi^{\left(  i\right)  }=\left(  \varphi^{\left(
i|u\right)  },\;\chi^{\left(  i|a\right)  }\right)  \,,\nonumber\\
& \left[  \varphi^{\left(  i|u\right)  }\right]  =r_{u}\,,\;\left[
\chi^{\left(  i|a\right)  }\right]  =s_{a}\,,\;\mathrm{rank\,}{\frac{\partial
\Phi}{\partial\eta}}=[\Phi]\,,\nonumber\\
& .i\leq u\leq k,\;1\leq i\leq k,\;i\leq a\leq k-1\,.\label{2.32a}%
\end{align}
The total Hamiltonian and and the Lagrange multipliers $\lambda$ have the
form
\begin{align}
& H^{\left(  1\right)  }=H+\lambda_{\varphi^{u}}\varphi^{\left(  1|u\right)
}+\lambda_{\chi^{a}}\chi^{\left(  1|a\right)  }\,\nonumber\\
& \,=H_{k}+\sum_{u=1}^{k}\Lambda_{u}\varphi^{\left(  1|u\right)  }+\sum
_{a=1}^{k-1}\lambda_{\chi^{a}}\chi^{\left(  1|a\right)  }\,,\nonumber\\
& H_{l}=H_{l-1}+\bar{\lambda}_{l}\varphi^{\left(  1|l\right)  }=H+\sum
_{u=1}^{l}\bar{\lambda}_{u}\varphi^{\left(  1|u\right)  }%
\,,\;l=1,...,k\,,\nonumber\\
& \Lambda_{u}=\lambda_{\varphi^{u}}-\bar{\lambda}_{u}\,,\;\;\bar{\lambda}%
_{u}=-\left[  M^{\left(  u\right)  }\right]  ^{-1}\left\{  \varphi^{\left(
u|u\right)  },H_{u-1}+\varepsilon\right\}  ,\nonumber\\
& M^{\left(  u\right)  }=\left\{  \varphi^{\left(  u|u\right)  }%
,\varphi^{\left(  1|u\right)  }\right\}  \,,\;\;\left(  H_{0}\equiv H\right)
\,.\label{2.33}%
\end{align}
The mutual commutation relations between the constraints are
\begin{align}
& u<v:\;\left\{  \varphi^{\left(  i|u\right)  },\varphi^{\left(  j|v\right)
}\right\}  =\left\{
\begin{array}
[c]{l}%
O\left(  \Phi^{\left(  1,...,i+j-1\right)  }\right)  ,\;i+j\leq u+1\\
O\left(  \Phi^{\left(  1,...,u\right)  }\right)  ,\;i+j>u+1,\;j\leq u\\
O\left(  \Phi^{\left(  1,...,j\right)  }\right)  ,\;j>u,
\end{array}
\right. \nonumber\\
& \left\{  \varphi^{\left(  i|u\right)  },\varphi^{\left(  j|u\right)
}\right\}  =\left\{
\begin{array}
[c]{l}%
O\left(  \Phi^{\left(  1,...,i+j-1\right)  }\right)  ,\;i+j\leq u+1\\
\left(  -1\right)  ^{u-i}M^{\left(  u\right)  }+O\left(  \Phi^{\left(
1,...,u\right)  }\right)  ,\;i+j=u+1\\
O\left(  \Phi^{\left(  1,...,j\right)  }\right)  ,\;i+\;j>u+1,
\end{array}
\right. \nonumber
\end{align}
and
\begin{align}
& \left\{  \varphi^{\left(  i|u\right)  },\chi^{\left(  j|a\right)  }\right\}
=\left\{
\begin{array}
[c]{l}%
O\left(  \Phi^{\left(  1,...,i+j-1\right)  }\right)  ,\;i+j\leq u+1\\
O\left(  \Phi^{\left(  1,...,u\right)  }\right)  ,\;i+j>u+1,\;j\leq u\\
O\left(  \Phi^{\left(  1,...,j\right)  }\right)  ,\;j>u,
\end{array}
\right. \nonumber\\
& \left\{  \chi^{\left(  i|a\right)  },\chi^{\left(  j|b\right)  }\right\}
=\left\{
\begin{array}
[c]{l}%
O\left(  \Phi^{\left(  1,...,i+j-1\right)  }\right)  ,\;i+j<k+1\\
O\left(  \Phi^{\left(  1,...,k\right)  }\right)  ,\;i+j\geq k+1,
\end{array}
\right. \nonumber\\
& \left\{  \varphi^{\left(  i|u\right)  },H_{j}+\varepsilon\right\}  =\left\{
\varphi^{\left(  i|u\right)  },H^{\left(  1\right)  }+\varepsilon\right\}
+O\left(  \Phi^{\left(  1,...,i\right)  }\right) \nonumber\\
& \,=\varphi^{\left(  i+1|u\right)  }+O\left(  \Phi^{\left(  1,...,i\right)
}\right)  ,\;j\geq i,\;u>i\,,\nonumber\\
& \left\{  \varphi^{\left(  i|i\right)  },H_{j}+\varepsilon\right\}  =O\left(
\Phi^{\left(  1,...,i\right)  }\right)  ,\;j\geq i\,,\nonumber\\
& \left\{  \chi^{\left(  i|a\right)  },H_{j}+\varepsilon\right\}  =\left\{
\chi^{\left(  i|a\right)  },H^{\left(  1\right)  }+\varepsilon\right\}
+O\left(  \Phi^{\left(  1,...,i\right)  }\right) \nonumber\\
& \,=\chi^{\left(  i+1|a\right)  }+O\left(  \Phi^{\left(  1,...,i\right)
}\right)  ,\;i<\;a\leq k-2\,,\nonumber\\
& \left\{  \chi^{\left(  i|i\right)  },H_{j}+\varepsilon\right\}  =\left\{
\chi^{\left(  i|i\right)  },H^{\left(  1\right)  }+\varepsilon\right\}
+O\left(  \Phi^{\left(  1,...,i\right)  }\right) \nonumber\\
& \,=O\left(  \Phi^{\left(  1,...,i\right)  },\varphi^{\left(  i+1|i+1\right)
}\right)  ,\;i\leq k-1\,.\label{2.34}%
\end{align}

The Poisson brackets between SCC from different chains vanish on the
constraint surface. The Lagrange multipliers $\lambda_{\chi}$ are not
determined by DP (and by the complete set of equations of motion). Whenever
FCC (SCC) exist, the corresponding primary FCC (SCC) do exist.

We note that imposing more restrictions on the theories under consideration,
we can obtain a more detailed constraint structure. For example, if we suppose
the rank constancy for the matrices $\left\{  \Phi^{\left(  1,...,i\right)
},\Phi^{\left(  1,...,i\right)  }\right\}  $, then we can conclude, at the
same $i$-th stage, that the constraints $\chi^{\left(  i-1|a\right)  }$ are of
the first class. However, the less restrictive condition of the rank constancy
for the matrix $\left\{  \Phi^{\left(  1\right)  },\Phi^{\left(  i\right)
}\right\}  $ is already sufficient for DP to be applicable. There exist some
models that obey the latter conditions only.

It is important to stress that commutation relations (\ref{2.34}) for $\chi$
mean that the property of a constraint to be or not to be of the first class
can be established only after completing the Dirac procedure. As a
consequence, in the general case, it is impossible to find special variables
(see \cite{GitTy90}) such that the first-class constraints have the canonical
form , i.e., it is impossible that FCC be some canonical momenta classified
according to the stages of the Dirac procedure. We consider an example of the
theory that confirms this statement. The corresponding Lagrange function is
\begin{equation}
L=\frac{1}{2}\left[  \left(  \dot{y}^{i}+x^{i}\right)  ^{2}+\left(
\dot{z}^{i}+y^{i}+E_{jk}^{i}x^{j}y^{k}\right)  ^{2}\right]
\,,\;i,j,k=1,2\,,\label{2}%
\end{equation}
where $E_{jk}^{i}$ is a constant matrix, which is antisymmetric with respect
to the lower indices$.$ The total Hamiltonian and primary constraints are
\begin{align}
& H^{\left(  1\right)  }=\frac{1}{2}\left[  p_{y^{i}}^{2}+p_{z^{i}}%
^{2}\right]  -x^{i}p_{y^{i}}-y^{i}p_{z^{i}}-E_{jk}^{i}x^{j}y^{k}p_{z^{i}%
}+\lambda^{i}\Phi_{i}^{\left(  1\right)  }\,,\nonumber\\
& \Phi_{i}^{\left(  1\right)  }=p_{x^{i}}=0\,.\label{3}%
\end{align}
We find that the second- and third-stage constraints are
\[
\Phi_{i}^{\left(  2\right)  }=p_{y^{i}}+E_{ik}^{j}y^{k}p_{z^{j}}%
=0\,,\;\Phi_{i}^{\left(  3\right)  }=p_{z^{i}}\,=0\,.
\]
All the constraints $\left(  \Phi^{\left(  1\right)  },\Phi^{\left(  2\right)
},\Phi^{\left(  3\right)  }\right)  $ are of the first class. In this case,
all $\lambda$'s remain undetermined and new constraints do not arise. The
commutator between the second-stage constraints,
\begin{equation}
\{\Phi_{i}^{\left(  2\right)  },\Phi_{j}^{\left(  2\right)  }\}=2E_{ij}%
^{k}\Phi_{k}^{\left(  3\right)  },\,\label{5}%
\end{equation}
is proportional to the third-stage constraints. We see that the second-stage
constraints are of the first class only with respect to the complete set of
constraints $\left(  \Phi^{\left(  1\right)  },\Phi^{\left(  2\right)  }%
,\Phi^{\left(  3\right)  }\right)  $ and not of the first class with respect
to the constraints $\left(  \Phi^{\left(  1\right)  },\Phi^{\left(  2\right)
}\right)  $ of the two first stages. Any constraint reorganization, which
respects the decomposition of the constraints according to the stages of the
Dirac procedure can not change this situation. The existence of special
variables, in which all $\Phi^{\left(  1\right)  }$ and $\Phi^{\left(
2\right)  }$ are canonical momenta, thus contradicts to relation (\ref{5}).

\section*{Acknowledgments}

The authors are thankful to the following foundations: Gitman to FAPESP, CNPq,
and DAAD; Tyutin to the European Community Commission (INTAS), grant 00-00262,
and to Russian Foundation for Basic Research, grants 99-02-17916, 00-15-96566.

\end{document}